\documentclass[aps,prl,twocolumn,superscriptaddress,groupedaddress]{revtex4}
\usepackage{subfig}
\usepackage{graphicx}  
\usepackage{dcolumn}   
\usepackage{bm}        
\usepackage{amssymb}   
\usepackage{slashed}
\usepackage{graphicx}				
\usepackage{amsmath}
\usepackage{mathtools}
\usepackage{tikz,pgf}
\usepackage{comment}
\usepackage{asymptote}
\usetikzlibrary{arrows,backgrounds}
\usetikzlibrary{fit,scopes,calc,matrix,positioning,decorations.pathmorphing}
\usepackage[all]{xy}
\usepackage{yfonts}

\newcommand{\Bracket}[1]{\ensuremath{\left\langle#1\right\rangle}}

\DeclareFontFamily{OMS}{oasy}{\skewchar\font48 }
\DeclareFontShape{OMS}{oasy}{m}{n}{%
         <-5.5> oasy5     <5.5-6.5> oasy6
      <6.5-7.5> oasy7     <7.5-8.5> oasy8
      <8.5-9.5> oasy9     <9.5->  oasy10
      }{}
\DeclareFontShape{OMS}{oasy}{b}{n}{%
       <-6> oabsy5
      <6-8> oabsy7
      <8->  oabsy10
      }{}
\DeclareSymbolFont{oasy}{OMS}{oasy}{m}{n}
\SetSymbolFont{oasy}{bold}{OMS}{oasy}{b}{n}

\DeclareMathSymbol{\smallleftarrow}     {\mathrel}{oasy}{"20}
\DeclareMathSymbol{\smallrightarrow}    {\mathrel}{oasy}{"21}
\DeclareMathSymbol{\smallleftrightarrow}{\mathrel}{oasy}{"24}

\begin{document}
\title{Coordinated inference, Holographic neural networks, and quantum error correction }
\author{Andrei T. Patrascu}
\address{ELI-NP, Horia Hulubei National Institute for R\&D in Physics and Nuclear Engineering, 30 Reactorului St, Bucharest-Magurele, 077125, Romania}
\begin{abstract}
Coordinated inference problems are being introduced as a basis for a neural network representation of the locality problem in the holographic bulk. It is argued that a type of problem originating in the "prisoners and hats" dilemma involves non-local signaling that is also found in the AdS/CFT duality. Neural networks are shown to have a significant role in the connection between the bulk and the boundary, being capable of inferring sufficient information capable of explaining the pre-arrangement of observables in the bulk that would lead to non-local precursor operators in the boundary.
\end{abstract}
\maketitle
\section{Introduction}
The holographic principle brings a radically new idea of how information can be encoded. Indeed, its name comes from optical holography, where a three-dimensional object emerges from the interference patterns of a two-dimensional holographic film. Moreover, it has long been known in mathematics that in order to characterise an object, one may use not only the object by itself, but also all the ways in which it can be mapped into other objects. Information about our initial object can be found both in the end result of the mappings, as well as in the maps themselves. Indeed, given some data or some algebraic object, a classifying space of that data corresponds to a space such that the maps from our object to that classifying space correspond to information about our original object (or data). The holographic principle implies that information about our bulk space can exactly be recovered in the boundary space. The mapping between the two is considered highly complex and indeed, the essence of holography. The holographic space has been derived by means of tensor networks in the bulk [1], [2] and various approaches to spacetime emergence have been considered [3]. The entanglement of the tensor networks was implicit. There exists a set of theorems linking the entanglement entropy to an extremal area within the bulk [4], [5], as well as certain approaches to describing the entanglement entropy by means of tensor networks encoding entanglement in the bulk [6]. 
However, it is important to understand that the map linking the two regions itself can provide additional information in a dynamical way. Speculations on this subject can be found in [7] but a better understanding of the methodology would be welcomed. 
The main goal of this article is to show that the way in which information from the bulk is mapped into the boundary also encodes parts of the information about the bulk in the good tradition of Grothendieck's relative point of view where information about an object $X$ is obtained by analyzing all morphisms $f:X\rightarrow S$ with $S$ some fixed object. This strategy shifts the focus from the objects of a category to the maps, providing a map-type description of the objects. For that, an object $x$ in a category can be described through the maps linking the object $x$ to some terminal object $term(C)$ of that category. 
The special type of map that can be used to recover bulk information on the boundary is a neural network. This neural network, by means of the Kernel-deep-network similarity [10] produces a superposition of training models encoded in the parameters of the network. This superposition, combined with the resulting global signaling technique produces a state of the network that cannot be reduced to any subsection of it, a phenomenon we call entanglement (which is not too different from the quantum mechanical entanglement). 

The application of neural networks to the holographic construction can be reduced (in a simplified form) to the prisoners and hats dilemma, a mathematical signaling problem that has been solved by means of a learning algorithm. The main aspect of this solution is the ability of a neural network to reduce the amount of information that has to be signaled and to predict the operator on the boundary before information propagating at the speed of light, originating in the bulk, can reach that boundary.

To make this point more explicit, I underline that entanglement can be seen from a categorial point of view as the obstruction to cartesianity of a monoidal category. Basically, the Set category of sets and functions between them is a cartesian monoidal category. The Hilb category of Hilbert spaces with linear maps between them is a non-cartesian monoidal category. In physics terms, this means that if we combine two systems each with their specific state spaces associated to them, we form a composite system that cannot be reduced to the cartesian pairing of the two subsystems. The tensorial product between the two spaces required in the non-cartesian category generates a larger composite space than the pairing between the two subspaces. The states that belong to that larger composite space but do not belong to any of the split subspaces are called entangled states. This is a phenomenon first encountered in quantum mechanics. However, there exist various functors relating the Hilb category to other categories, one example was the Cobordism category. There too can a form of entanglement be defined. However, what is of interest here is the connection between the Hilb category and a supposed category of neural networks. As neural networks are constructed in such a way that the parameters encode a linear superposition of all training data, we have to admit that they form a category functorially connected to Hilb. By this, they will have an equivalent of entanglement, that will be used as such in what follows. It is important to underline that this entanglement is mathematically the same as the entanglement we find in quantum mechanics, and hence deserves the same name.

Indeed the process of introducing global information on the boundary (for example in the form of Wilson loops as precursors [8]) requires a series of bulk phenomena to take place that are not easily explainable via any other non-neural tools. An event in the bulk will have to naturally propagate information about its occurrence towards the exterior, reaching the boundary and causing certain effects there. However, if the holographic principle is to hold exactly, the information about the bulk space must already be encoded in the boundary, although maybe in a non-local fashion, before the signal from the bulk can reach the boundary. Neural networks are capable of explaining the two phenomena introduced by Polchinski, Toumbas, and Susskind in [13] in order to account for such a situation. To describe the effects, I provide the reader with an exact quotation from that article : 
\begin{quotation}

"At this time the entire boundary suddenly “lights up” with a spherically symmetric distribution of energy which in total equals the mass of the black hole. In other words a fraction of the energy originally stored in the collapsing shells very quickly flows and is redistributed into a homogeneous component. We will call this phenomenon “light-up”. This behaviour seems extremely bizarre. The instantaneous re-arrangement of energy appears to violate causality. However this may not be so. To better understand it, we will describe an analogous example involving a sudden flow of electric charge. Consider an example in which initially there is a concentration of charge in some region $R_1$. At time $t=0$ the charge is found to disappear from $R_1$ and reappear at $R_2$ which is outside the forward light-cone of $R_1$. To see how this can happen, imagine a wire connecting $R_1$ and $R_2$.The wire is full of electrons and positive ions so that it is electrically neutral. Now we prearrange observers at each point of the wire so that at $t=0$ they move each electron slightly toward $R_2$. The result is a sudden appearance of charge at the ends with no charge density ever occurring anywhere else. If there was already a charge at $R_1$ it would be cancelled by the new charge at that point. The net result would be a sudden redistribution of charge. Two ingredients are necessary for such behaviour.  The first is that the current vector $j_\mu$ be spacelike. Since the charge density on the wire is always zero it is clear that the current is purely in the space-like direction. This also means that in some frame the charge density was negative. This of course is not a difficulty since charge density can be either positive or negative. The other ingredient is prearrangement. The physical conditions along the wire must include agents with synchronised clocks that are instructed in advance to act simultaneously. The sudden flow of energy requires the same two ingredients. In order that the energy is rearranged so suddenly, the flux of energy $\Bracket{T^{0i}}$ must be much larger than the energy density itself.  This means that the energy density can be made negative by a Lorentz boost. However, unlike electric charge, energy is not allowed to be negative in SYM theory. Since in classical SYM theory the energy density is positive this kind of flow of energy is absolutely forbidden in the classical theory. However quantum theory allows local negative energy densities as long as they are (over)compensated by nearby positive energy density"
\end{quotation}
Of course, this is a simplified example. The connection may not be 1-dimensional as it is described there, and the more modern approaches consider tensor networks to describe the proper entanglement in the bulk. However, one remains with the problem of pre-arrangement which implies some form of learning that must occur at some point. For this to happen naturally, we need some technique of spreading global information that will not be recoverable by means of any local interaction. To learn the signaling technique for this we do need some learning process, that is probably best described via neural networks. In fact, the prisoners and hats problem provides a method of "guessing" the right configuration and the exact type of action (or response) by each one of the agents involved, with only one possibly wrong guess, and that with probability of 50 percent.

\section{Prisoner's dilemma}
There exists a set of problems called coordinated inference problems that have been recently solved via machine learning and neural networks and that I argue, play a role in understanding how the holographic maps function. Here too, global information is spread via local signals resulting in a global state that contains more information than what can be retrieved locally. The procedure of generating this is obtained via machine learning techniques. The main idea emerged from a game introduced to mathematicians by [12]. The problem is stated as follows: a set of prisoners is placed in a row, each facing in the direction of his next fellow in the same direction. A set consisting of blue and red hats is being placed on each of the prisoners' heads without them knowing their respective colour. They will only see the following colours and they can hear the previous guesses. With only one prisoner allowed to guess wrongly, what is the strategy that insures each of the others will guess correctly? The solution is to use the first prisoner to signal to the rest the parity of the red (or blue) hats he sees. He looks only to the prisoners in his front, and hence receives information only from one direction. This suggests a causal structure, however, interestingly enough, there are guessing strategies developed in [12] where this "causal structure" is further limited, leading to what we would call in physics a horizon. 
Whenever the next prisoner who is being asked, will see a change of parity, he will know his hat changed that parity and hence each of them, except the first one, will be able to guess his own colour. This problem implies several aspects relevant to the bulk-boundary map in AdS/CFT and to the connection between neural networks and quantum information.  The way of thinking is as follows: the parity being a piece of global information about the hats, is used to improve the statistics of guessing for the missing (invisible) local colour (that of the prisoner having to guess his own hat colour). One bit (first guess) will be used therefore to improve the overall probability of guessing by generating a piece of global information shared across the observers. 
The prisoners and hats problem stated as it is, is static, and of course, could not, by itself be a model for the holographic principle. However, the procedure through which a neural network solves that problem could be, as it is the inference that defines holography, and not the problem itself. The ability of a neural network to infer from the data a signaling prescription that generates entanglement is what is significant for the holographic map. A neural network can be trained to do this and the results have been obtained in [15]. The idea is to train the agents to make the proper guesses, resulting in the encoding of the bulk information in the non-local observables of the boundary.

This has various consequences also in the opposite direction, as we can imagine a method of extracting information across a gauge-string duality by having local operators ("letters" and "words" using the Polyakov terminology [23]) being inferred using global information encoded in a string theory. That information can be reconstructed by means of Wilson lines/loops and would correspond to a bulk neural network capable of inferring them via learnable procedures.

An important part of this procedure is the signaling aspect that conveys global information. The state describing the prisoners by parity is not separable, being an analogue of an entangled state, having global information that cannot be retrieved locally. While the signaling itself occurs in a causal manner, the global information must be part of the entanglement wedge. Interestingly, this example connects entanglement and the global information it conveys with local transmission of information, resulting in a non-local boundary observable which represents the whole information about the row of prisoners. Therefore, the neural network can develop a signaling procedure that in its turn allows for the proper inference (guessing) of a quantum boundary operator. What is even more fascinating is that this problem has been solved via classical machine learning, the solution being a valid strategy for approaching the problem (a guessing strategy). Therefore, a neural network was capable of identifying a global solution to this problem and predicting a quantum (precursor) operator by using a configuration that did not use any of the usually employed quantum circuits. In this case the parity is a global property of our 2-colour group. Other types of global information can be used and hence a more advanced solution can infer more information about the chain of prisoners if, for example, group homology is employed.


The role of the neural network was to find the specific inference methodology required for the efficient transmission of information such that the boundary observables appear to be precursor-like (non-local). Basically the observation here is that in order to get from local bulk observables to non-local Wilson loops on the boundary, the holographic map needs a neural network component or some other method of advanced optimization or deep learning capable to generate a similar result. 
The ability of neural networks to link local and global information has a fundamental role in understanding many-particle strongly coupled systems and is currently regarded as a mystery: how is it possible that strongly coupled highly entangled quantum systems can so well be described by means of classical neural networks? Indeed, if we take into account one property of entanglement, namely that the global information obtained by putting together subsystems into a global system, cannot be split back into the two subsystems, then we may notice that a similar property is at work in neural networks, particularly in those solving multi-agent optimization problems as for the inference problem stated above. The prisoners and hats problem is a 1-dimensional problem, hence it is, if anything, simply a starting point for a new approach. One should not forget that other techniques designed to solve such problems, like, say the DMRG are working precisely only for 1-dimensional problems. The neural network approach however could be generalised to higher dimensions, although the example presented in this introduction is only 1-dimensional. 


Let us have a look at how neural networks designed to solve such problems would work. First, the neural network would optimise for a multi-agent problem in which each agent can develop an action policy designed to optimise some global loss function. Therefore, at each layer of the neural network one advances by passing linear data from all the input edges of a neuron, through a non-linear activation function. The non-linear role of the activation function is basically just to eliminate a negative signal and to cap and normalize the transmitted signal to unity or to the value of the linear combination obtained from the integration of the input signals. In the next neuron layer, another linear transformation depending on the parameters of the next layer will be introduced, and so on, until the information is combined together on each layer, and passed towards the output. However, at each such layer, a backpropagation effect will have to take place, which is basically our learning process. This will depend on a more global loss function that is designed to produce the global strategy of our agents. Through this, our nodes will gain access to some levels of global information about our problem and enforce, layer by layer, an optimization of the parameters. Due to the non-linear nature of the activation functions, when this globalization of information is being performed, parts of the local information won't be available anymore. Moreover, the global information will be able to better describe each node, according to our global loss function and our strategy, than the local information being passed forward could ever do. What we therefore obtain is an entangled state. We will gain some form of global information in our neural network, that can never be recovered by looking at any specific neuron or neuron layer in our network. While the fact that neural networks have the ability of performing tasks that can predict global information better than any algorithmic approach focused only on the local data is known, noticing that this is precisely the effect we obtain when we construct an entangled wavefunction is rather new. Because of this, not only can a classical neural network perform better in problems like those describing strongly coupled many-particle systems, but it can in principle also give us global information on the boundary, information that could not be obtained by studying only the local propagation of the information from the bulk. 

\section{Neural networks, global information, and entanglement}

To prove this in more detail, let us look at how path integral quantisation works, and how, in some key aspects, it brings in elements that can be recovered in neural networks with multiple layers. 
Let us therefore look at what it means to describe a process involving interactions in terms of Feynman's path integrals, and how that same mathematical underpinning is achieved by neural networks, in a somewhat different but equivalent way. The mathematical term for such an analogy is functoriality. Ineed, it seems that if at some point we could define a categorial theory of neural networks, it would have some functorial relation to the path integral quantisation prescription a la Feynman. 
This can be seen also from the kernel-deep-learning similarity described in [10], [11]. There it is shown that the parameters of the network encode linear combinations of the training examples presented to them. Indeed, the learning process allows the construction of a specific kernel that then can be used as a filter for the problem at hand. But in this sense the learning process of the neural network looks very similar to what quantisation does, both procedures being defined essentially over the Hilb category and taking into account linear combinations at a level of maps between Hilbert space states. It may therefore be enlightening to show how the standard Feynman path integral quantisation works and how it is similar to the learning process of a neural network. 
Let us follow Feynman's approach and imagine an experiment consisting of three measurements of three observables, $A$, $B$, and $C$. There will be several possible outcomes for these measurements, and suppose we obtain answer $a$ from the measurement of $A$, then $b$ from the measurement of $B$ and finally $c$ from the measurement of $C$. Assume also that the measurements of $A$, $B$, and $C$ completely specify the state of the quantum system at hand. Let then $P_{ab}$ be the probability that if one measures $A$ and obtains $a$ then the measurement of $B$ will produce $b$. Similarly we define $P_{bc}$ and we can define $P_{abc}$ the probability that if $A$ gives $a$, then $B$ gives $b$ and $C$ gives $c$. Classically we would then write that 
\begin{equation}
P_{abc}=P_{ab}P_{bc}
\end{equation}
However, in quantum mechanics this would be true only if $B=b$ would be a complete specification of the state, which is usually false. 
If we write 
\begin{equation}
P_{ac}=\sum_{b}P_{abc}
\end{equation}
we assume that there must have been some intermediate value $b$ occurring when we passed from $A$ to $C$ and that happened with a probability $P_{abc}$. We sum over all mutually exclusive alternatives of $b$. In quantum mechanics this is wrong! Instead, we define three complex numbers 
$\phi_{ab}$, $\phi_{bc}$ and $\phi_{ac}$ such that 
\begin{equation}
\begin{array}{ccc}
P_{ab}=|\phi_{ab}|^{2}, & P_{bc}=|\phi_{bc}|^{2}, & P_{ac}^{q}=|\phi_{ac}|^{2}\\
\end{array}
\end{equation}
and we replace the classical law of composition with 
\begin{equation}
\phi_{ac}=\sum_{b}\phi_{ab}\phi_{bc}
\end{equation}
The classical mistake made here is to assume that there was a well defined state $b$ through which our system had to go to reach $C$. Indeed, it was not. Such a state does not exist unless one defines a specific observable to determine it, and if one only measures $A$ and $C$, the intermediate state is not, truly, defined. 
This is the first step to constructing the phase structure of the wavefunction and to make entanglement possible, as a global, locally irreducible construction. This is something that emerges in neural networks most naturally. First, a neural network will integrate in one of its layers all the information obtained from the previous layers, and pass it through a series of linear transformations, with various weights. But then one passes this through an activation function to reach the next layer, inducing a series of probabilistic threshold functions that will create a set of improperly realized intermediate states. The neural network therefore will have the same ability to work with various internal states, without making any of them more realistic or actually occurring. The name "hidden layers" is suggestive for that. Indeed, the neural network will also be able to couple with the intermediate states by means of the backpropagation in the same way in which the path integral approach does, as will be shown in what follows. That is why, as in the probability interference pattern we see in quantum mechanics, we will have a similar interference occurring in the neural networks. Also, in the same way in which in quantum mechanics, if we attempt to verify the reality of the intermediate steps, we will interfere with the proper evolution of the experiment and change the outcomes, in neural networks, we cannot simply break the layers of the network after the optimization or learning process started without altering the results and making them invalid.  In fact the ability of the neural networks to integrate intermediate values of possible outcomes, without directly measuring them is what the complex phase is doing in the path integral formulation. The deeper and broader the neural network, the more such paths can be integrated and then optimised accordingly. 
To see this, one should look at the definition of paths in the three situations: the classical probability of reaching point $B$ starting from point $A$, the path integral quantisation probability of doing the same, and the neural network connection between the input $A$ and the output $B$. In this way it will be clear that the neural network and the Feynman path integral approach to quantisation have several important points in common. It is important to note that in the case of the Feynman path integral approach we don't just sum over all intermediate paths, as we would when we would describe the situation in classical physics. We have to take into account the complex phase nature of the states involved, and their ability to interfere in any region of space $R$ one considers, and it is this that the neural network paths have in common with the quantisation prescription. There too, we allow the paths to interfere in every region $R$ defined by a layer of neurons, allowing the signal to pass to the next layer by means of an activation function. The neural network learning process is the one that can actually generate entangled states in the neural network, providing a global link between subsystems that cannot be recovered globally. To see this let us have a look at how Feynman interpreted his probability for an amplitude of a space-time path. Even on a 1-dimensional problem the relevance of the complex amplitudes becomes relevant. If we consider a 1-dimensional path and a particle that can take up values for the corresponding coordinate $x$ at different time moments separated by an interval $\epsilon$ then we can attach a set of pairs of measurements $x_{1}, ..., x_{n}$ to times $t_{1},...,t_{n}$, with $t_{i+1}=t_{i}+\epsilon$. In a classical context a path is defined by the succession of values $x_{1}, ... x_{n}$ taken by the coordinate. The probability of such a path is given by $P(..., x_{i}, x_{i+1},...)$. To find the classical probability of the path being in a certain region $R$ of spacetime we have to integrate $P$ over that region. This would be wrong in quantum mechanics, unless one specifically measures and determines all intermediate point values $X_{i}$. If one does not fundamentally possess such a knowledge, then one goes to the complex amplitudes with their imaginary parts and asks what possible paths one could have obtained in that region $R$, the answer being that in order to obtain the probability that our particle is in the region $R$ one has to integrate $|\phi(R)|^{2}$ over the region $R$, and to this $\phi(R)$ one has to include all the histories of the paths that can end up in $R$ hence
\begin{equation}
\phi(R)=\lim_{\epsilon\rightarrow 0} \int_{R}\Phi(...,x_{i},x_{i+1},...)...dx_{i} dx_{i+1}...
\end{equation}
$\Phi$ depending on all the variables that define the whole path and hence being the probability amplitude function of the paths $x(t)$. As Feynman noted, "if an ideal measurement is performed to determine whether a particle has a path lying in a region of space-time, then the probability that the result will be affirmative is the absolute square of a sum of complex contributions, one from each path in the region."
The next step relies on calculating the quantity $\Phi(R)$. The paths contribute equally in magnitude, but their respective phases depend on the classical action, as we well know when we perform the path integral. This is the time integral over the Lagrangian taken along the path. 

Now let us see what a neural network is doing in general. The state of a neural network is defined by the parameters associated to each of the linear transformations required to pass from one neuron to the next or from layer to layer. These are generated dynamically and usually optimised by some backpropagation of a gradient. Independent of the configuration one takes, a neural network will propagate information and integrate it via repeated applications of linear transformations. Therefore one cannot truly trace one single path from the input to the output inside the network. What one can do is to trace the probability of each signal to be found in some region of the network, and that probability depends on all the collateral signals that have been integrated in it via the multiple applications of various linear maps. Therefore, at each neuron, the output depends on all the possible paths of the adjacent signals that are in the region where that specific neuron has connections. That is basically our region $R$. At each level there is a non-linear threshold function that creates a gap in the signal, where an adjacent one can contribute and shift the whole path, which gives the trajectory the probabilistic nature required, namely that structure that integrates all possible neighbouring paths of the region $R$. As in quantum mechanics, there is no defined intermediate state that one could associate to any potential output. This is why those layers are called "hidden". If we attempted to identify the inputs and outputs at those levels, we would distort the normal working of the neural network. Of course, there is a fundamental difference. The neural network remains classical, hence, at least in principle we could identify a signal in the hidden layers. Such a signal would however be mostly meaningless for any outside observer. To obtain a truly quantum neural network one would have to replace the links between neuron layers with quantum circuits. However, the classical neural network comes the closest to having the mapping structure of a quantum entangled system. It is up to us to decide how to interpret the output result. If we want to interpret it classically, the optimization process will lead us to a classical result that we simply pick out at the output. This approach however will mask a series of outcomes that would become visible if we decided to interpret the result as quantum, apply a Born-type operation, and integrate the end result. In that case, our outcome will be more quantum and will encode parts of the global structure of our problem. The back-propagation generates some form of entanglement, as it brings back information on the global result, by means of a Loss function defined over all the possible actions taken by agents, and hence about the global structure of our problem's manifold, and makes the network non-separable, or at least, if we decided to cut the network and use only pieces of it after the integration of the global optimization data, we would not be able to obtain the same global information. 
This hopefully clarifies how a neural network behaves like a quantum system, although it does not start as a quantum system by itself, and not every link in the network is a quantum circuit in the usual sense.

Let us look at how this type of problem can be linked to the bulk operators in AdS/CFT. 

\section{precursor operators and kernel methods}
To see how this problem relates to a holographic scenario let us look at the thought experiment performed in ref. [13]. An interesting situation occurs in the context of a collision of two wave packets emitted from two distinct points $x_{1}$ and $x_{2}$ at time $t=-\pi/2$. From a boundary perspective the behaviour of the system after $t=0$ seems extremely sensitive to the details of the emission process. Let there be two packets of fixed energy emitted from diametrically opposite points such that in the boundary theory the process starts as a pair of expanding thin energy shells. The shells will meet at $t=0$. One may expect that after $t=0$ the evolution to be extremely sensitive to the initial conditions. This has been proven false in [13]. Indeed, the two shells will hardly interact because as $R\rightarrow \infty$ the points near the boundary become widely separated from their sources and the gravitational field equations linearise. Our energy momentum remains expressed on the thin shells contracting towards their opposite points. In this context therefore $\Bracket{T_{\mu\nu}}$ is zero everywhere off the shells together with all the other boundary fields corresponding to classical supergravity in the bulk. If we accept that $\Bracket{T_{\mu\nu}}=0$ classically, then we must think at a vacuum-like classical local state for the system. All functionals of the fields supported off those thin shells should be vacuum like. However, in quantum theory this cannot be true, as the region between the shells will have to be excited away from the local vacuum configuration despite the zero expectation value for the energy density. Let the two packets collide head on at $t=r=0$ as in the example of ref. [13]. I must follow reference [13] carefully at this point so that the analogy with the neural network approach to the coordinated inference problem becomes evident. If the energy of the colliding particles is of the order of the Planck mass, the result will be a Schwarzschild black hole. Some of the energy of the reaction will be transferred to the black hole and this will become spherically symmetric and start emitting Hawking radiation quickly enough. The entire history of the black hole lasts a finite amount of time which tends to zero like $N^{-1/4}$ in our large-N theory. The fields at the AdS boundary cannot respond to this until the light from the event has propagated from $r=0$ to the boundary at the time $t=\pi/2$. The resulting signal will pass the boundary in a time $\delta t\sim (Ng^{2})^{-1/4}$ and that is when the entire boundary lights up in a spherically symmetric way encoding the entire history of the black hole. Such an instantaneous re-arrangement of degrees of freedom seems to violate causality. However, as argued in ref. [13] this must not be so. If we look back at the quotation from [13] we notice that the main problem is the pre-arrangement of the observers in the bulk. The example is simpler because we are dealing with only a 1-dimensional problem. However, the neural network required to deal with this problem does not change in its structure if we expand the problem to higher dimensions. Of course, one may require more neurons and layers, and the depth of the network may have to change, but the overall idea of how the network solves the signaling problem remains the same. 

This may seem an implausible example but it may be exactly what happens if we re-interpret the experiment above in terms of our coordinated inference problems. Indeed, generating the strategy of guessing the colour of the hat is similar to designing the strategy involving coordinated, pre-arranged observers. This strategy can be found via neural networks. It implies the identification of the global information to be signalled as being precisely the instruction of when to measure. This information can be provided in a causally connected manner, resulting in quantum entanglement across the observers and hence in the emergence of a  "pre-arrangement". Can such a strategy be generated spontaneously? It seems that the answer is positive. 
This example is indeed 1-dimensional, but it can be generalised without difficulty to higher dimensions, at the cost of a more complex neural network. 

First one has to understand what entanglement means. Given two spaces containing the states of two systems, say space $H_{1}$ containing the states of system $X_{1}$ and space $H_{2}$ containing the states of system $X_{2}$, if one considers the unitary linear transformations allowed in quantum mechanics one notes that combining the systems $X_{1}$ and $X_{2}$ doesn't result in a space of states like the classical space, namely the cartesian product $H_{1}\times H_{2}$ but instead we obtain the tensorial product of the spaces $H_{1}\otimes H_{2}$. The distinction is extremely important in terms of information, as the resulting space contains global information about the system that cannot be retrieved by looking only at each subsystem in part. The prisoners and hats problem puts into evidence precisely this structure, while noting that the global information is encoded, by means of the learning phase, in the original signaling strategy developed by our neural network. In that sense, our neural network has inadvertently (or not) learned how to entangle our system of prisoners with global information that it developed in the learning phase. In what sense is the information not recoverable locally? Without the learning phase, if we take each prisoner separately and interrogate it about the colour of its hat, the result will have a 0.5 probability of success. The learning phase of our neural network produced by an interference of individual "paths" taken by the agents leads to a result that lets all prisoners give the precise answer every time. Basically our neural network, in its learning phase, produced an entangled global system. What we do with complex phases in the path integral quantisation, we do with the learning phase and the associated memory in the neural network. Here too we deal with a space of probability amplitudes that are not having a specific realisation but that allow the integration of neighbouring signals via the input links of a neural layer. 

There are some general assumptions that I challenge here. First, it is wrongly assumed that the ability of producing an entangled state and of maintaining it as such is directly linked to Planck's constant $\hbar$ and hence is somehow "extremely small". While, certainly a quantum effect, entanglement can however be visible over large scales, at any energy, and may even have cosmological implications. It is not $\hbar$ that determines the entanglement structure, but instead the logic behind the algorithm we create. Entanglement is simply global information that cannot be recovered locally in the subsystems of our larger system. It is the result of the fact that the maps in the $Hilb$ category are linear maps, and not structureless pairings as is the case in the $Set$ category, and therefore, we need to construct our composite spaces of states via the tensor product and not via the cartesian product, a procedure that creates a lot of global information that remains inaccessible to the subsystems taken separately [26]. Of course, at the most fundamental level, $\hbar\neq 0$ determines that outcome. But its smallness has no implication on the scale at which the entanglement can be witnessed.  It has often been noticed that entanglement can occur over any distance scale, and efforts are being done to keep it active at higher temperatures. 
Exactly the same procedure takes place in a neural network. There too, we cannot in principle form a "cartesian" neural network that can be separated at will, without altering its computational capabilities. Indeed, it has been learned that the topology of a neural network can be decisive in solving certain types of problems. 
I imagine that the prisoners and hats problem will become a benchmark problem of how a neural network can learn a signaling procedure that encodes global information, helping local agents make a globally informed decision. This type of problems can be solved taking into account the linear maps required in a neural network to advance signals or backpropagate gradients.

Basically the arguments above bring us to the conclusion that there exists a functorial relation between neural networks (defined in a categorial sense) and the category of Hilbert spaces $Hilb$. 
Let us consider the $Hilb$ category [9]. By definition it is a concrete dagger category that has a tensor product $\otimes$ and a tensor unit $I$ such that $(Hilb_{\mathbb{C}}, \otimes, I)$ is a symmetric monoidal dagger category, and $I$ is a simple object and a monoidal separator. It also has a biproduct $\oplus$ and a zero object such that $(Hilb_{\mathbb{C}}, \oplus, 0)$ is a semiadditive dagger category. We also need a dagger equaliser such that $(Hilb_{\mathbb{C}}, \oplus, 0, eq)$ is a finitely complete dagger category, a direct colimit $lim_{\rightarrow}(F)$ for any inductive system $F$, and a morphism $g:B\rightarrow C$ for any dagger morphism $f:A\rightarrow B$ such that $g=eq(f, 0_{B}\circ 0^{\dagger}_{A})$. 
The question would be, whether a neural network can be constructed in a similar way? The answer seems to be yes [10], [11]. A functor would map a Hilbert space as an object of our initial category into a neuron or a layer of neurons, where the pairing between them would be defined by the linear maps that relate elements of the Hilbert space, and where the tensoring between Hilbert spaces would be mapped into tensoring the neurons, bringing them together in layers, or even connecting layers between each other. Each morphism between Hilbert spaces would therefore by definition be mapped into a similar morphism for the neural networks. This should be seen more like an analogy than a definitive proof, and it is certainly possible that the functor may not be full and faithful. 
However most of the algebraic structures found on the $Hilb$ side are to be detected on the neural network side. This means that from a functional point of view, the two categories act in a similar way. That doesn't make neural networks completely quantum, as the choice of interpreting the maps on neural networks as linking wavefunctions is strictly subjective and depends on the use of some form of Born rule mechanism at the end, however, from all other points of view, the two systems act the same. 

\section{The neural network contribution}

With this in mind, the neural network will work in a two-phase process: the first will be the learning process, involving the optimization of the network inside the bulk, and the second will be the operational phase, in which the model will be applied to generate the output on the boundary of our AdS space. The resulting model will act as a code for the non-local operator generated on the boundary. Of course the information regarding parity that has been considered in the first example is only limited, but there is enough additional group topological information available to insure the inference of more complex information associated to each insertion (represented in our example by the type of hat colour that needed to be guessed). Discussions on how one can expand the hat colour problem to situations in which the distribution of colour is continuous or there is an infinite countable or uncountable number of prisoners have been presented in [12].

It is important to point out that the idea that neural networks can be used to derive the AdS/CFT duality has been discussed in reference [14]. Here I show that neural networks play a foundational role in the way a signal in the boundary is retrieved in the bulk before any locally propagating impulse would have the time to reach the boundary. Entanglement plays a fundamental role, but its structure is represented by the non-separable global information distributed to the observers by means of the strategy developed by the neural network.

 Reinforcement learning is usually understood by imagining an agent interacting with an environment and trying to maximise a reward function given to him. This often arises as a Markov decision process (MDP) which is characterised by a set of states, S, a set of actions, A, the probability of state transition $P_{a}$, and the reward given for executing an action $a_{t}$ at a given state $s_{t}$. If stochasticity is absent one can set $P_{a}=1$. Given a policy $\pi$, the goal is to construct an agent that realizes that policy $\pi(s_{t})=a_{t}$ in such a way that the expected reward $E[\sum_{t=0}^{\infty}R(s_{t},a_{t})|\pi]$ is maximised. In this way the policy is slowly learned and that requires usually for the future reward to be discounted by a parameter $\gamma$. The policy can be learn if the reinforced learning algorithm is constructed via neural networks parametrised by certain weights and biases $\theta$. Also, the Q-value defined as 
\begin{equation}
Q(s_{t},a_{t})=r_{t}+max_{a_{t+1}}Q(s_{t+1},a_{t+1})
\end{equation}
being the numerical estimation of the reward following the agent taking action $a_{t}$ at state $s_{t}$, and $r_{t}$ being the reward and $max_{a_{t+1}}Q(s_{t+1},a_{t+1})$ the maximum future value of the reward. The Q-function is represented by a neural network. The learning policy is defined for discrete actions spaces and can be formulated as the policy parametrised by $\theta$ to choose the action that has a maximum Q-value. The policy can be learned by neural networks using the Bellman equation in order to calculate the mean squared loss function and then the gradients needed for back-propagation
\begin{equation}
L_{t}(\theta)=E[(r_{t}+max_{a'}Q(s',a';\theta)-Q(s,a;\theta))^{2}]
\end{equation}
This is the standard DQN algorithm. However, such an algorithm often faces convergence problems because the max operation can lead to over-estimations of the $Q$ value. Improvements involve the usage of a separate target network for predicting the future $Q$ value inside the max operation, or using dueling DQNs which have separate network heads that predict the advantage and value components of the Q-value, distributional DQNs, noisy nets, and others.

The essential aspect of this proposal is the learning phase which implies adapting a geometry to an entanglement structure on one side, and learning a signaling strategy that would convey the desired entanglement. Of course, this can be applied also in the opposite way, deriving/learning what type of entanglement corresponds to a given geometry. One could also try to allow the network full freedom to seek both the global signaling/entanglement structure and the geometry, given the type of boundary field theory we wish to encode as well as other boundary conditions on the field structure.

The automated learning phase is capable of independently discovering new ways in which data can be represented. However, reference [10] beautifully shows that in fact networks learned by the gradient descent algorithm are equivalent to kernel machines. The kernel appears to be used as a similarity function and allows the prediction of the new representations. From this perspective the network weights appear as nothing but a superposition of training examples. Indeed, the network incorporates information regarding the target function into the kernel, which will encode ultimately our global information. These two observations, namely that the kernel methods are similar to the neural network approaches [10] and the fact that the kernel includes superpositions of learned examples incorporated in the neural network, brings allows us to directly employ neural network strategies to derive the signaling method capable of encoding the information from the bulk in a non-local manner in the boundary. In fact, all our holographic constructions are kernel based and hence amenable to such an approach, while the global information applied on such a superposition will create a global inseparable state (entanglement) within our network. 

As the addition of an integral of a local operator to the conformal action is a way of perturbing the conformal field theory, one can consider the addition of the square of an operator $\mathcal{O}$ dual to a fundamental field in gravity. This would amount to a double trace deformation. The dual operation of perturbing the conformal field theory is to change the boundary conditions imposed on the bulk fields. In the case of a scalar operator computed with CFT methods the expectation value for the perturbed CFT will look like
\begin{equation}
\Bracket{\mathcal{Q}}_{f}=\Bracket{Q e^{-\frac{f}{2}\int d^{d}x\mathcal{O}^{2}(x)}}_{CFT}
\end{equation}
Using a Hubbard-Stratonovich auxiliary field in the large $N$ limit one can obtain the two point function
\begin{equation}
\Bracket{O(k)O^{\dagger}(q)}_{f}=(2\pi)^{d}\delta(k-q)\frac{1}{f+k^{d-2\Delta}}
\end{equation}
where $\Delta$ is the conformal dimension of the operator $\mathcal{O}$. 
For $\Delta<d/2$ this correlation function can be interpreted from a renormalisation group point of view. It shows how the relevant operator $\mathcal{O}^{2}$ added to the conformal action generates a renormalisation group flow away from the conformal point leading to another conformal point where the operator has dimension $d-\Delta>d/2$. On the gravity side we obtain the same physics but starting with the correlation function obtained from the generating functional 
\begin{equation}
W[J,f]=log\Bracket{e^{\int d^{d}x[-\frac{f}{2}\mathcal{O}^{2}+J(x)\mathcal{O}(x)]}}_{CFT}
\end{equation}
where $f\mathcal{O}^{2}$ is the perturbation and $J\mathcal{O}$ is the source. Appropriate boundary conditions arise from the variational principle if the corresponding boundary terms are added to the action. The connection between the two can be determined by an optimization network specific to neural networks.

To make this clearer let us think in terms of gauge-string duality. In the strong coupling of a lattice gauge theory elementary excitations can be seen as closed strings made up of colour-electric fluxes. If quarks are present the closed flux tubes end up splitting and ending on quarks suggesting quark confinement. In the $SU(N)$ gauge theory the string interaction at large $N$ is weak. We may expect that in the continuous limit the best description is that of flux lines and not of fields. This suggests a duality between gauge fields and strings. There is also a class of superstrings that contain quantum gravity. It could be possible to see those as flux lines of some unknown gauge theory. Hence we could regard spacetime only as a quasiclassical limit of some gauge theory in which spacetime is not yet emerging [23]. In this context the observables are sets of gauge invariant operators formed as products of some elementary ones. These have been historically called "letters", the observables we wish to obtain being "words". One is interested in the dynamical correspondence between strings and those "words". String theories have an infinite number of gauge symmetries in the target space, generated by the sequence of the zero norm states on the world-sheet. The lowest symmetry there is the general covariance, leading to the conservation of the energy-momentum tensor in the gauge theory. Higher gauge symmetries generate other relations between the words resulting from the equation of motion of the gauge theory [23]. Gauge-string duality claims that there is in fact an isomorphism between the gauge invariant operators and the vertex operators of certain closed string theories in the background. It is known that for the conformal group of the gauge theory side there is expected to be a group of motion for the metric on the string side. This leads to a spacetime with constant negative curvature which is the AdS spacetime. This has been derived via machine learning. The question in a more general case would be what kind of problem is to be solved to obtain a different realisation of this duality? Our gauge invariant observables (words) of the gauge theory have been represented as a set of agents guessing their local group element (the hat colour). The connection between them in the case of the prisoners and hats problem is the simple colour group I presented above. The global information can be signalled through a parity measure and produces a non-separable state at the level of the agents. That state allows us to infer information that cannot be accessible via local observables. This would be the equivalent of precursor operators on the boundary in the case of AdS/CFT. However in a more general setting one can consider the operators and implement a neural network optimising for the proper entanglement structure and then using it to infer the string theory side. This is how new dualities could emerge. However, this will be the subject of another paper. The question here remains: how can Nature figure out in the most effective way what is the pre-arrangement required for the observers to produce the information on the boundary? The proposal in this paper is that the bulk structure could play the role of a neural network trained to find the type of signaling that would allow the inference of non-local information. Indeed, this signaling has a classical transmission part, but as for any quantum protocol, a significant part of the information is available via entanglement. The entanglement can then be fed to the neural network trained to provide us with a geometry or the other way around, one can extract a signaling protocol by the method described for the derivation of the guessing method of the hat problem. The boundary conditions can of course be altered such that perturbations are introduced in the conformal field theory and the signaling method can be analysed via our neural network. However, in this article I take for granted the fact that neural networks can be used to either derive the AdS/CFT duality or to derive potentially new dualities as shown above. I try to suggest however something more profound, namely that a neural network approach is what is needed to insure the pre-arrangement needed for non-local precursors to arise on the boundary as encoders of local bulk information. The essential pre-arrangement presented in the example of [13] is provided by an inference tool generated by an (potentially quantum i.e. with nodes represented by quantum field creation/annihilation operators, see [22]) neural network in the bulk.

The neural network in the bulk space can be seen as a generalisation of the tensor network approach where it plays the fundamental role it has: a general optimizer. The aspect we considered to be curious in the thought experiment described above and in [13] was the existence of a pre-arrangement requirement that seemed to be at odds with any form of ordering that could emerge considering just causal connections. It also seemed to be at the foundation of the emergence of precursor operators (expected to be fundamentally non-local Wilson loops) on the boundary. The question asked was : is it possible to infer information from the bulk that doesn't have the required causal connection to the boundary? The answer seems to be positive. To exemplify, consider the single trace operator written as
\begin{equation}
\mathcal{O}_{i_{1},...,i_{n}}=Tr ( \Phi_{i_{1}}\cdot \Phi_{i_{2}}\cdot ... \cdot \Phi_{i_{n}})
\end{equation}
where $\Phi_{i}$ is the elementary field observable and the single trace operator is a matrix product of those defined in the adjoint (matrix) representation. The trace insures the gauge invariance. Such single trace operators correspond to single string excitations and they behave like "words" (strings of "letters" in the definition of Polyakov), the argument of the trace being mapped onto a superstring in spacetime. The idea has also been discussed in the so called string bit model [24] where the string is being regarded as a composite object similar to a long polymer of infinitesimal string bits, each of them being described by some dynamical variables encoding the position and the momentum. Such a model has various difficulties, although it could be amenable to a neural network approach starting with string vertex operators. Taking this model partially seriously, the construction in the bulk (or, equivalently on the string side of the gauge/string duality) corresponds to a set of bits in the form of a tensor network of links. Locality at this level has been properly defined. The inference of the type of boundary (or, more generally gauge theory) operators is not clear. It is clear that the entanglement structure is responsible for the bulk spacetime structure, but there remain issues related for example to the pre-arrangement of the operators in the bulk that appear to be somehow arbitrary. That pre-arrangement of operators is what can be formulated by using neural networks. The bulk network would be a neural network in which the discrete bits would form nodes of such a neural network. Allowing the learning process to change the links between the nodes would be equivalent to string couplings and vertices. But it is clear that such vertices are to be associated via duality to gauge theory operators. It is necessary that we also include multiple trace operators in order to properly encode AdS/CFT and the non-local precursors on the boundary. But we also know that AdS/CFT is only one realisation of the gauge/string duality. The process of inferring the strategy of the prisoners is analogue to the process of identifying a signaling and an entanglement structure that would lead to a certain type of colouring structure for the set of hats. The colouring of all the hats of the prisoners is the analogue of a Wilson line observable. The communication between them represents a causal structure which can be modified accordingly. Ref. [12] shows several situations in which signaling exists, is bidirectional, one-directional, or even restricted, all corresponding to various types of causal structures with or without horizons in the interpretation presented here. The colouring in the prisoners problem has global structure and it makes sense in the "flux-tube" section of the theory. The inference strategy and the entanglement plus signaling process makes sense on the gauge side of the theory where we deal with "letters and "words" i.e. single and multiple trace operators in a non-abelian gauge theory. Keeping the analogy in this way we see that neural networks are a key ingredient in the process of linking non-local observables in the gauge theory to local ones on the string side but the inference is not restricted to a specific emergent spacetime. 

 Machine learning experiments on specifically designed neural networks have been performed and the hat problem signaling has been successfully learned [15], [25]. Let us see how this could be applied to the holographic context and what each learning phase would ensue. 
Entanglement can be seen as global information that is not accessible locally. Splitting our state into parts will not convey anything about the information encoded by the entangled state. This is exactly the solution our machine learning algorithm must reach in the first place. It must "realize" that the first qubit is supposed to encode information about a global indicator (say parity). The type of network used was a deep distributed recurrent Q-network (short DDRQN) [16]. The goal there was to teach a neural network to develop its communication protocol, but it is interesting to note that the result was basically a quantum communication protocol as it created an entangled state via its first signaling. In physics we consider a semiclassical limit of large N and large 't Hooft coupling with free local fields in the bulk and we wish them to be encoded in the conformal field theory on the boundary. Given a radial coordinate $z$ that is equal to zero on the boundary and a boundary field $\phi_{0}(x)$ we will have 
\begin{equation}
\phi(z,x)\sim z^{\Delta}\phi_{0}(x)
\end{equation}
then the bulk field can be expressed in terms of the boundary field by means of a kernel $K$ as
\begin{equation}
\phi(z,x)=\int dx' K(x'|z,x)\phi_{0}(x')
\end{equation}
Our $\phi_{0}(x)$ corresponds to the local operator $\mathcal{O}(x)$ in the boundary CFT. Hence the local bulk fields are expected to be dual to non-local boundary operators 
\begin{equation}
\phi(z,x)\leftrightarrow \int dx' K(x'|z,x)\mathcal{O}(x')
\end{equation}
Our function $K(x'|z,x)$ also called "smearing function" basically defines our holographic map by which local bulk excitations are encoded on the boundary. These smearing functions are essential in understanding the causal structure of the bulk and hence are the main subject of learning for our neural network. The signaling solution found by our neural network will determine these functions and eventually generate the entanglement structure we need to convey our information on the boundary. 
Consider the $AdS_{3}$ with Rindler coordinates given the metric 
\begin{equation}
ds^{2}=-\frac{r^{2}-r_{+}^{2}}{R^{2}}dt^{2}+\frac{R^{2}}{r^{2}-r^{2}_{+}}dr^{2}+r^{2}d\phi^{2}
\end{equation}
with $-\infty < t,\phi < \infty$ and $r_{+} < r < \infty$. $R$ is our AdS radius and $r_{+}$ is the radial position of the Rindler horizon. Our bulk function can be expected to have the form 
\begin{equation}
\phi(t,r,\phi)=e^{-i\omega t}e^{i k \phi}f_{\omega k}(r)
\end{equation}
with 
\begin{widetext}
\begin{equation}
f_{\omega k}(r)=r^{-\Delta}(\frac{r^{2}-r_{+}^{2}}{r^{2}})^{-i\hat{\omega}/2}F(\frac{\Delta-i\hat{\omega}-i\hat{k}}{2},\frac{\Delta-i\hat{\omega}+i\hat{k}}{2},\Delta,\frac{r_{+}^{2}}{r^{2}})
\end{equation}
\end{widetext}
where $\hat{\omega}=\omega R^{2}/r_{+}$ and $\hat{k}=kR/r_{+}$. The mode functions $f_{\omega k}$ are real. 
We can expand in Rindler modes 
\begin{equation}
\phi(t,r,\phi)=\int_{-\infty}^{\infty}d\omega\int_{-\infty}^{\infty}dk a_{\omega k}e^{-i\omega t}e^{ik\phi}f_{\omega k}(r)
\end{equation}
The Rindler boundary field is given by 
\begin{equation}
\phi_{0}(t,\phi)=\int_{-\infty}^{+\infty}d\omega \int_{-\infty}^{+\infty}dk a_{\omega k}e^{-i\omega t}e^{ik\phi}
\end{equation}
with 
\begin{equation}
a_{\omega k}=\frac{1}{4\pi^{2}}\int dt d\phi e^{i \omega t} e^{-ik\phi}\phi_{0}(t,\phi)
\end{equation}
Hence the bulk field in terms of the boundary field is
\begin{widetext}
\begin{equation}
\phi(t,r,\phi)=\frac{1}{4\pi^{2}}\int d\omega dk (\int dt' d\phi'e^{-i\omega(t-t')}e^{ik(\phi-\phi')}\phi_{0}(t',\phi'))f_{\omega k}(r)
\end{equation}
\end{widetext}
The mode functions $f_{\omega k}$ diverge at large $k$ and hence we need to use an analytical continuation method to imaginary values of the $\phi$ coordinate [18]. 
In a quantized theory the excitations produced by the responsible creation operators must be adapted such that the process occurs in a pre-defined manner. Our neural network therefore keeps the characteristics of the neural network used in solving the prisoners and hats problem but its underlying structure is closer to the one introduced by G. Dvali in ref. [18]. What will be achieved is a set of operators producing the required excitations in the bulk in a manner learned by the neural network during the training phase. The bulk space will therefore host not only the quantum error correction protocols alone, capable of correcting a certain number of errors, but also the learning phase of the neural network by means of a dynamical allocation of the links. This dynamical allocation of the network links will not only generate a non-trivial topological structure and hence the required entanglement but will also generate the bulk geometry. 
To better understand this let us have a look at the neural network designed to solve the prisoners and hats problem. The first level of complexity would imply "single agent fully observable reinforcement learning". An agent is capable of observing its current state at each discrete time step and choses an action according to a decision policy, observes the rewards, and then transitions to a new state. The expectation of the return $R_{t}$ is the function it wishes to maximise: 
\begin{equation}
R_{t}=r_{t}+\gamma r_{t+1}+\gamma^{2}r_{t+2}+...
\end{equation}
where $\gamma$ is a discount factor. Given a policy $\pi$, a current state at time $t$, $s_{t}$, and an action at time $t$, $a_{t}$ the $Q$ function is the expectation of the return function
\begin{equation}
Q^{\pi}(s,a)=\mathbb{E}[R_{t}|s_{t}=s, a_{t}=a]
\end{equation}
We have an optimal action value function which obeys the Bellman optimality equation
\begin{equation}
Q^{*}(s,a)=\mathbb{E}_{s'}[r+\gamma \max_{a'}Q^{*}(s',a')|s,a]
\end{equation}
Deep Q-networks (DQN) are usually parametrised by $\theta$ and are optimised by the optimization problem associated with the loss function for each iteration $i$
\begin{equation}
L_{i}(\theta_{i})=\mathbb{E}_{s,a,r,s'}[(y_{i}^{DQN}-Q(s,a;\theta_{i}))^{2}]
\end{equation}
with a target 
\begin{equation}
y_{i}^{DQN}=r+\gamma \max_{a'}Q(s',a';\theta_{i}^{-})
\end{equation}
where $\theta_{i}^{-}$ are weights of a target network frozen for a number of iteration.
The second level of complexity would imply a multi-agent setting in which each agent can observe the states of all the other agents and selects an individual action according to a team reward shared among all agents. This is particularly important as it introduces the global parameters required for learning the signaling solution of the hat problem and in our case it is the first step towards implementing the quantum entanglement that must result in the final output. However, in order to introduce the causal structure and our spacelike current we need to give up on full observability. Some (or all) of the states of the other participants will be hidden from the current observer and his decision must be based only on a limited observation of the other participants and the correlations it observes with the actual states of the other participants.

Such a correlation while capable of giving some information is not capable of disambiguating the other hidden states themselves. This aspect is the second stage required for entanglement construction and/or learning. The observation correlated with the participant $s_{t}$ is denoted $o_{t}$. The neural network will be a deep recurrent Q-network in which $Q(o,a)$ is approximated with a recurrent neural network that can maintain an internal state and aggregate observations at each time step. Therefore our network can learn an entanglement structure and it can use entanglement if provided to it, in order to learn a geometry. Partial availability of information and correlation of nodal states implies a tracing at the level of each agent leading to generation of entanglement between them. 

Returning to our Rindler-AdS situation in order to analytically continue the coordinate $\phi$ we perform the Wick rotation $\tilde{\phi}=i\phi$. This brings us to a de-Sitter geometry 
\begin{equation}
ds^{2}=-\frac{r^{2}-r_{+}^{2}}{R^{2}}dt^{2}+\frac{R^{2}}{r^{2}-r_{+}^{2}}dr^{2}-r^{2}d\tilde{\phi}^{2}
\end{equation}
where $r$ is a time coordinate. We periodically identify $\tilde{\phi}\sim \tilde{\phi}+2\pi R/r_{+}$. 
The de Sitter invariant distance function is
\begin{equation}
\sigma=\frac{rr'}{r_{+}^{2}}(cos(\frac{r_{+}(\tilde{\phi}-\tilde{\phi'})}{R})-\sqrt{1-\frac{r_{+}^{2}}{r^{2}}}cosh(\frac{r_{+}(t-t')}{R^{2}}))
\end{equation}
In the bulk space the field can be expressed in terms of the retarded Green's function which is the imaginary part of the commutator inside the past light-cone of the future point, while it vanishes outside this region. The bulk field therefore can be written as
\begin{widetext}
\begin{equation}
\phi(r,\tilde{\phi},t)=\int d\tilde{\phi}'dt'\frac{r'(r'^{2}-r^{2}_{+})}{R^{2}}G_{ret}(r',\tilde{\phi}',t';r,\tilde{\phi},t)\tensor{\partial_{r'}}\phi(r',\tilde{\phi}',t')
\end{equation}
\end{widetext}
The region of integration is over a spacelike surface of fixed $r'$ inside the past light cone of the current point.
This is directly representable in terms of our tensor network. Indeed the information required for the neural network to make predictions of the type involved in the pre-arrangement of observers implies integrating over the partial information obtained from the rest of the network by means of correlations available to the current agent describing the states of the others. As those must be spacelike separated as is the case in the problem of observers making measurements in a pre-arranged fashion, we have to conclude that the integration over the spacelike surface of fixed $r'$ implies the reconstruction of global information for the use of our neural network. This corresponds to a more general type of neural network, one that is susceptible of solving the types of problems we need. Particularly we need "multi-agent partially observable reinforcement learning" and the most profitable method implies deep distributed recurrent Q-networks.

Each agent is provided with its previous action as input to its next time step. The agents explore the optimization space stochastically and hence this method mixes the integrated actions of the surrounding nodes with the past trajectories of each agent. It is precisely this type of integration that provides the non-locality required to generate the proper boundary operators. However, this remains insufficient considering the type of optimization needed. The learning phase must employ some form of weight sharing across the agents. This is provided by both the integration and by the retarded Green function. However, due to the incomplete knowledge of the surrounding states it is unlikely for two agents to receive the same input data from the rest of the network, given the types of correlations involved, allowing each of them to behave differently, generating each their own internal states.

Another requirement is to decide what can be done about experience replay. This translates in our setting by allowing the agents to learn independently their environment. In the solution to the hat problem this option is disabled as this would lead to each agent learning a potentially different environment as the network will appear non-stationary to each of them. In fact, this option should be considered when a quantum approach is taken as this would amount to a path integral quantisation over all possible intermediate states. This approach would enormously increase the complexity of an actual computation but it has strong theoretical support if what we wish to describe is a quantum theory in the boundary. In the limit in which $r'\rightarrow \infty$ the retarded Green function becomes using ref. [19], [20]
\begin{widetext}
\begin{equation}
G_{ret}\sim i(c(-\sigma-i\epsilon)^{-1+i\sqrt{m^{2}R^{2}-1}}+c^{*}(-\sigma-i\epsilon)^{-1-i\sqrt{m^{2}R^{2}-1}}-c.c)
\end{equation}
\end{widetext}
with 
\begin{equation}
c=\frac{\Gamma(2i\sqrt{m^{2}-1})\Gamma(1-i\sqrt{m^{2}-1})}{2^{2-i\sqrt{m^{2}-1}}R\Gamma(\frac{1}{2}+i\sqrt{m^{2}-1})}
\end{equation}
In our prisoners and hats problem the global information signalled by the first bit of information must have a global nature, hence it encodes properties eventually encoded in the group (co)homology of our colouring. The branch cuts have been chosen along the positive real $\sigma$ axis allowing for a single valued expression of the multi-patched surface. The resulting globally non-trivial structure is represented through the entangling information signalled by the first observer. Of course, the choice of the branch cut itself is arbitrary and its structure may reveal additional global properties. Our neural network will learn to make different choices in this aspect and find the best signaling/entangling parameter to be sent. It would be interesting to see how the situation changes when there can be more signaling bits or maybe when the fist signaling state is a qubit.

This procedure protects against errors in guessing the colouring structure. Of course parity is not protected against different colour picks, but that is not the point of the method. The protection is against making wrong guesses, and the errors in guessing are permanently protected by the identification of the parity of the colours as given. The predictors we construct, be it from parity or other forms of global indexes (potentially tensors of arbitrary rank) are protecting against guessing errors, and not against some erroneous flip of a colour. 
 Of course one has to accept a maximum of the guessing strength or else global information would perfectly encode and/or be encoded by local information which we know from the basic idea of entanglement to not be the case. A similar bound on the amount of quantum information a strategy can protect against guessing errors results from the holographic principle itself, which makes the holographic principle seem like a limitation in extracting local information from global geometry. Given the holographic map, in our case the kernel/smearing function, it is obvious that the holographic principle appears as an obstruction to the existence of local boundary degrees of freedom emerging from bulk operators. A local operator in the bulk is mapped to a non-local extended operator on the boundary. The holographic principle basically expresses the obstruction for the bulk to boundary continuation of the number of degrees of freedom in order for the extension to non-local operators in the boundary to make sense. 

\section{inference, prisoners and holography}
In what follows I will explain in parallel, showing the connections between the hat guessing strategy inference and the inference of the measurements required in the holographic context with a neural network in the bulk. The main aspect of the solution is to find a strategy on which the prisoners can agree upon when performing the guessing game. The same is valid for the observers in our holographic context. We consider that the formation of the black hole and its evaporation is a strictly local event, hence there is causal "room" for a strategy learning. By the way a neural network operates, there will be entangled states formed, namely states that can have a global interpretation without such interpretation being retrievable locally in the subsystems separately. To express it as a multi-agent reinforced learning problem, let's consider the state space $s=(s^{1}, ... , s^{n}, a^{1}, ... , a^{n})$ with $n$ the total number of agents and $s$ the space of possible outcomes with $s^{m}=\{blue, red\}$ the colour of the $m$-th hat, while $a^{m}=\{blue, red\}$ is the action taken by the agent. This has an equivalent in the decision of the agent to act and perform a measurement in the given holographic context. On the $m$-th step, the agent $m$'s observation is 
\begin{equation}
o^{m}=(a^{1}, ..., a^{m-1}, s^{m+1}, ..., s^{n})
\end{equation}
We are constructing the learning phase for the agents via our neural network, providing rewards starting with an overall reward zero and increasing it as our agents manage to obtain a globally correct answer. The repetition created by the neural network approach is basically obtained in the holographic problem by means of a quantum circuit, making full use of the bulk space ability of performing wavefunction engineering, a feature that is yet to be achieved practically. However, as shown before, entanglement as global information that cannot be obtained from the separated pieces of the system is there, both in the neural network model and in the holographic model. The reward is given at the end of the episode when the number of agents guessing correctly is given as a reward: $\sum_{m}(\delta(a^{m},s^{m}))$. In the holographic context this would amount to having all observers making the pre-required measurements to produce the global information on the boundary, hence, it would have to encode the fraction of the local bulk information recovered in the boundary precursor. The quantum nature of the holographic context would allow the information to be stored as a probability amplitude and the result to be inferred automatically. The classical neural network would do something similar, with the difference that, of course, each branch of the network would transmit a classical signal so the training phase would involve several repetitions. Once the initial, causally connected learning phase has been completed, the optimised parameters can be used to infer the signaling procedure and to allow for a pre-arrangement of the observers in both contexts. The first observer is hit by the shockwave and starts signaling a pre-established code containing global information to the rest of the prisoners. That signal will travel at the same rate as our shockwave hence, causally, but will encode global information that will never be accessible without an inference policy deduced via the initial use of a neural network. The entangled state will be there whether we use a classical or a quantum neural network. However, it will be clearly quicker to do everything with a quantum neural network, as the various learning possibilities would be carried out simultaneously and the relevant superposition of states would emerge quicker. It is assumed that Nature would have access to a quantum network. With the signaling procedure obtained by the network, the observers will be able to provide the precise measurements that will encode the bulk information in a non-local Wilson loop observable. There is one difference: in the prisoners and hats problem the one prisoner will always determine the parity by looking at the others, hence the global information is retrieved via two ways communication. In the case of holography, the information is established in the learning phase, as the phenomena occurring in the bulk are easily proved to be local and causally connected. Aside of that, the processes of inference are similar and this difference doesn't pose a fundamental problem: global information is recovered causally when the event occurs in the bulk, or via local two-ways communication in the case of the prisoners. As always, entanglement needs some initial causal connection.

 \section{conclusion}
What may not be clear yet is that solutions to coordinated inference problems can be regarded as methods to improve the guessing of the naively unavailable information. We can define the general framework for hat problems as follows: let there be a set $A$ of agents, a set $K$ of colours, and a set $C$ of functions (colourings) mapping $A$ to $K$. The agents wish to construct a coordinated strategy such that if to each agent is given some information about one of the colourings (hence a global information) then he can provide a guess related to local aspects of the colouring (his guess about his own hat colour). The collection of guesses taken for all agents in the set is capable of restricting the colourings to the one real colouring existing in the given context (a random choice from all possible colourings). Those guesses must be consistent for all agents in the set. The information provided to an agent $a\in A$ is given by an equivalence relation on the set of colourings indicating that agent $a$ cannot distinguish between the two colourings found to be equivalent by that relation. As can be seen, our tool implies erasure of portions of the knowledge about the colouring, as it is invisible to an agent given his context. That information is being "erased" from his domain of accessibility.

A guessing strategy, usually involving some form of connection between global and local data, implies recovering the lost information. We can define a guessing strategy according to [12] as a map $G_{a}$ for agent $a$ from $C$ to the power set of $C$, namely $P(C)$ such that if the guessed colouring and the actual colouring are equivalent for agent $a$ then $G_{a}(f)=G_{a}(g)$. That means that agent $a$ is guessing that colouring $f$ belongs to the set $G_{a}(f)$ of colourings, with the goal that the guessed set becomes minimised and converges towards the actual colouring. This process, involving a predictor $P(f)=\cap\{G_{a}(f):a\in A\}$ implies the restoration of the local data. In more general contexts we may define a visibility graph allowing agents to see only parts of their environment, say one way visibility, etc. Even our initial simple example implies the recovery of the agents own hat, i.e. extraction of a local point with some parts of the global information. This is therefore indeed a form of error correction. 
In this article I showed several new and important aspects: first and as main subject, I showed that coordinated inference problems play a fundamental role in the study of locality in the bulk and its relation to non-locality on the boundary. Indeed, I linked a neural network approach capable of generating a coordinated strategy for the solution of the basic hat problem with the pre-arrangement of observers in the bulk capable of generating non-local precursors in the boundary. Second, I understood that such inference problems are error correction problems and that there is a limit to error correction for any code as suggested by holography. Moreover, I connected the idea that such a limit to the amount of error correction emerges from a form of sheaf cohomology seen as an obstruction to linking local information in the bulk to non-local spacetime subsets in the boundary. 
Future work would imply a more thorough formulation of the holographic principle in terms of sheaf cohomology and the implementation of a NEAT algorithm that would generate both geometry and topology in the bulk given a global signaling/entanglement structure.
As a conclusion, in this article it was shown that the pre-arrangement of observers required by ref. [24] in order to obtain the information from the bulk on the boundary can be obtained via a learning phase if one assumes a neural network acting in the bulk. The optimization would require some form of superposition of potential strategies (also represented as paths) and the final global information signalled during the procedure would lead us to an inseparable state at the level of the observers. 
Therefore, it is possible to train a network to re-construct the bulk information in a non-local Wilson-loop type form on the boundary, without having a direct causal link between the two. All processes of learning within the bulk remain of course strictly causal.

\end{document}